\documentclass[12pt]{article}
\usepackage{amsxtra,amssymb,amsthm,amsmath,latexsym}

\theoremstyle{plain}

\newtheorem{theorem}{Theorem}[section]

\newtheorem{lemma}{Lemma}[section]

\newtheorem{remark}{Remark}[section]

\def\ds{\displaystyle}
\def\nd{\noindent}

\def\R{{\mathbb R}}

\def\oH{\buildrel\circ\over H}
\def\oH1{\buildrel\circ\over H\kern-.02in{}^1}
\def\qed{{\hfill $\Box$}}

\def\bysame{\rule{.5in}{.005in},\ }

\begin{document}


\title{
A non-overdetermined inverse problem of finding the potential
from the spectral function
   \thanks{key words: Non-overdetermined inverse problem,
    spectral function, Schr\"odinger equation  }
   \thanks{Math subject classification: 35R30 }
}

\author{
A.G. Ramm\\
 Mathematics Department, Kansas State University, \\
 Manhattan, KS 66506-2602, USA\\
ramm@math.ksu.edu\\
}

\date{}

\maketitle\thispagestyle{empty}

\begin{abstract}
Let $D\subset \R^3$, be a bounded domain with a
$C^{\infty}$
connected boundary
$S$, $L=-\nabla^2+q(x)$ be a selfadjoint operator defined in
$H=L^2(D)$ by the Neumann boundary condition,
$\theta(x,y,\lambda)$ be its spectral function,
$\theta(x,y,\lambda):=\ds\sum_{\lambda_j<\lambda} \varphi_j(x)\varphi_j(y)$,
where $L\varphi_j=\lambda_j\varphi_j$,
$\varphi_{j N}|_S=0,$
$\|\varphi_j\|_{L^2(D)}=1$,
$j=1,2,\dots$.
The potential $q(x)$ is a real-valued function,
$q\in C^\infty(D)$.
It is proved that $q(x)$ is uniquely determined by the data
$\theta(s,s,\lambda)\ \forall s\in S$,
$\forall \lambda\in \R_+$ if all the eigenvalues of $L$ are simple.
\end{abstract}


\section{Introduction}

Let $D \subset \R^n $, be a bounded domain with a
$C^{\infty}$
connected boundary
$S$,
that is, locally the equation
$x_n = f(x^\prime)$, $x^\prime := (x_1 \dots x_{n-1})$,
of $S$ is given by the function $f(x^\prime) \in C^{\infty}$.
We assume $n=3$. This assumption is used only in the proof of
Lemma 3.1 below and can be dropped. If $n>3$ 
one should refer to the existence of the coordinate
system in which the metric tensor, used in
the proof of lemma 3.1, has the property: $g^{in}=0$ for
$1\leq i <n$. The basic ideas of our proof, outlined below,
are valid for $n\geq 3$.

Let $L= -\Delta + q(x)$, where $\Delta$ is the Laplacian and
$q(x) \in C^\infty (D)$, the potential, is a real-valued
function.
Let
$\varphi_j (x), \quad j=1,2, \dots,$
be the normalized real-valued eigenfunctions of the 
selfadjoint operator $L$ defined
in $H=L^2(D)$ by the Neumann boundary condition:

\begin{equation}
L \varphi_j = \lambda_j \varphi_j \hbox{\ in\ } D,
\quad \| \varphi_j \| = 1,
\quad \| \varphi \| :=
\| \varphi \|_{L^2(D)},
\tag{1.1} \end{equation}

\begin{equation}
\varphi_{jN} = 0 \hbox{\ on\ } S,
\tag{1.2}\end{equation}
where $N$ is the unit normal to $S$ pointing into $D^\prime:= \R^n
\backslash D$.
We choose $\varphi_j$ to be real-valued functions. By elliptic regularity,
the functions $\varphi_j(x)$ are $C^{\infty}(D)$ and
$C^{\infty}(S)$ on the boundary $S$. 

The spectral function of $L$ is defined by the formula: 
\begin{equation}
\theta (x,y, \lambda):=
\sum_{\lambda_j < \lambda} \varphi_j (x) \varphi_j (y),
\tag{1.3} \end{equation}
the eigenvalues $\lambda_j$ are counted according to their multiplicities.

The inverse problem (IP) we are interested in is:

IP: {\it given $\theta (s,s,\lambda) \quad \forall s \in S$ and
$\forall \lambda \in \R_+$, find $q(x)$.}

We are concerned with the 
{\it uniqueness of the solution to IP}. This inverse
problem 
{\it is not overdetermined}: the data is a function of $n$ variables
$s, \lambda$\, ( $S$ is an $(n-1)$--dimensional manifold,
$\lambda \in
\R$) and
the unknown $q(x)$ is also a function of $n$ variables.

Most of the non-overdetermined inverse problems are {\it open}
\cite{R1}--\cite{R2}, for example:

a) inverse potential scattering problems with the data
$A(\alpha^\prime, \alpha_0, k)$ (fixed direction 
$\alpha_0$ of the incident wave, all
directions $\alpha$ of the scattered wave and all energies $k>0$), 

b) inverse backscattering problem with the data
$A(-\alpha, \alpha, k)$, 

c) inverse obstacle
scattering problem with the data $A(\alpha^\prime, \alpha_0, k_0)$
(fixed incident direction, fixed frequency and all directions of the scattered
wave), 

and

d) inverse geophysical problem with the data $u(x^\prime, x_n = 0,k)$
(surface data for all frequencies, fixed position of the point source on the
surface).

It seems that no uniqueness theorems for multidimensional inverse
scattering-type problems
with non-overdetermined data have been obtained so far.
The IP, studied in this paper, is a multidimensional one of the above type
and with non-overdetermined data. 
 We prove a uniqueness theorem for this problem.

The inverse problem with the overdetermined data, which is the
 spectral function $\theta (x,y,\lambda),$ 
known $\forall x, y \in S$ and $\forall \lambda \in \R$
has been considered in \cite{Ber1}--\cite{Ber3}, \cite{An},
\cite{R1}, \cite{R3}-\cite{R5} .

In \cite{R1}, \cite{R3}, \cite{NSU} (see also Remark 2.1 below) 
the following uniqueness theorem is obtained:

\begin{theorem}
The data
$\{\lambda_j, \varphi_j (s)\}_{\substack{\forall s \in S\\ \forall j}}$
determine
$q(x)$ uniquely.
\end{theorem}

The result we want to prove is:

\begin{theorem}
{\it The IP has at most one solution if all the eigenvalues of $L$
in (1.1)--(1.2)  are simple}.
\end{theorem}

 It is possible that the conclusion of Theorem 1.2 holds without the
assumption concerning the simplicity of the eigenvalues, but we do not
have a proof of this currently. Generically
one expects the eigenvalues to be simple, if there are no symmetries in
the problem.

The strategy of our proof is outlined in the following three steps:

\nd {\it Step 1.} \ \ $ \theta (s,s,\lambda) \Rightarrow \{\lambda_j,
\varphi_j^2 (s) \}_{\substack{\forall s \in S\\ \forall j}}$

{\it This step is done under the additional assumption that all the
eigenvalues  $\lambda_j$ are simple.}

It is an open question whether Theorem 1.2 holds without this
assumption.
Any assumption that will make Step 1 possible is
sufficient for our purpose.
 
\nd {\it Step 2.}\ \  $ \varphi_j^2 (s) \Rightarrow \varphi_j (s)$

\nd {\it Step 3.}\ \  Apply Theorem 1.1 to the data
$\{ \lambda_j, \varphi_j (s) \}_{\substack{\forall s \in S\\ \forall j}}$

Step 1 does not require much work.

Step 2  requires the basic work.

 Let us outline the ideas needed for
the completion of Step 2. 

One can prove that $\varphi_j (s)$ cannot 
have zeros of infinite order,
that is, if $y \in S$ and
$|\varphi_j (s)| \leq C_m|s-y|^m \quad \forall m = 1,2, \dots ,$ where
$C_m$ are positive constants and $s \in S$ are arbitrary points
on $S$, then $\varphi_j(x) \equiv 0$. In the Appendix it is
proved that if a real-valued function $f\in C^{\infty}(S)$ does
not have zeros of infinite order on $S$, 
then, up to its sign, the
function $f$ is uniquely defined on $S$ by its square $f^2$.
A procedure for finding,  up to its 
sign, 
$\varphi_j(s)$ on $S$ from the knowledge of $\varphi_j(s)^2$
on $S$ is described below.

It is known that $\varphi_j (s)$ cannot vanish 
on an open (in $S$) subset of
$S$ (uniqueness of the solution to the Cauchy problem for elliptic
equations). Thus, $\varphi_j(s) \neq 0$ for almost all points of $S$.
 Take an arbitrary point $s_0$ at which
$\varphi_j^2 (s_0) \neq 0$. By the continuity 
of $\varphi_j^2 (s)$ on $S$,
there is a maximal domain ${\mathcal F} \subset S$
containing the point $s_0$ in which $\varphi_j^2(s) \neq 0$.
Define $\varphi_j(s) := \sqrt{\varphi^2_j (s)} > 0 \hbox{\ in\ }
{\mathcal F}$.
Let ${\mathcal L} \subset S$ be the boundary of 
${\mathcal F}$. We prove that
$\varphi_j(s)$ is uniquely determined in ${\mathcal F}^\prime :=
S \backslash {\mathcal F}$.
The problem
is to determine the sign of $\varphi_j (s)$ for 
$s \in {\mathcal F}^\prime$
in a neighborhood of ${\mathcal L}$, where 
$\varphi_j (s)=0$ when $s\in {\mathcal L}$.

The basic idea for determining this sign is to extend 
$\varphi_j (s)=
\sqrt{\varphi^2_j(s)}$ from $\mathcal{F}$
across  $\mathcal L$ with maximal smoothness. Such an extension
is unique and determines the sign of $\varphi_j (s)$ outside of 
$\mathcal F$. For example, if one defines $\sqrt{x^2}=x>0$ for
$x>0$, then the unique maximally smooth extension of $\sqrt{x^2}$
in the region $x<0$ is $\sqrt{x^2}=x$.

To determine the sign of $\varphi_j (s)$ outside of
$\mathcal F$, calculate the minimal integer $m$ for
$\varphi_j(s) = \sqrt{\varphi^2_j (s)}, 
\quad s \in {\mathcal F} \subset S$,
 such that
$ \gamma ({\overline s}) := 
lim_{s \to {\overline s}} \frac{\varphi_j (s)}
{|s - {\overline s}|^m} \neq 0$.
Here $s \to {\overline s} \in {\mathcal L}$ along 
a curve $\ell$ originating at a point in $\mathcal F$,
transversal to 
 ${\mathcal L}$, and passing through a point
${\overline s} \in {\mathcal L}$ into  
$\mathcal F^\prime$.

The integer $m< \infty$ does exist if the zeros of
$\varphi_j(s)$ are of finite order. That these
zeros are indeed of finite order is proved in
Lemma 3.1 in section 3.

Let us describe the way to continue $\varphi_j(s)$ along
the  curve $\ell$ across  ${\mathcal L}$ into  $\mathcal
F^\prime$.

Define
$\hbox{sgn\ } \varphi_j (s) = (-1)^m$, where
$s \in {\mathcal F}^\prime \subset S$ is any point in a  
sufficiently small
neighborhood of $\overline{s}$. This algorithm determines uniquely
$\varphi_j (s)$ for all $s \in S,$ given the data $\varphi_j^2 (s)$.
Since the eigenvalues are assumed simple, the above algorithm
produces the trace $\varphi_j (s)$ of the eigenfunction
$\varphi_j (x)$ on $S$.

Since $q(x)$ and $S$ are $C^{\infty}$-smooth up to the
boundary, so are the eigenfunctions $\varphi_j (x)$, and
the above algorithm produces a $C^{\infty}$-smooth function on $S$.
Therefore one can use the Malgrange 
preparation theorem (\cite {CH}, p.43) to study
the set of singular points of $\mathcal L$.

The above argument deals with the continuation of an 
eigenfunction through
its "zero line" $\mathcal L$ on $S$.
There are at most finitely
many points on a compact surface
 $S$ at which several "zero lines" intersect
and only finitely many "zero lines" can intersect at one point.
Otherwise there would be a point on $S$ which is a 
zero of infinite order
of $\varphi_j (s)$, and this is impossible by Lemma 3.1 of section
3 below.

If one continues $\varphi_j (s)$ across $\mathcal L$ 
by the above rule, the
function $\varphi_j (s)$ will be uniquely determined on all of $S$
by the choice of its sign at the initial point $s_0$.

In the Appendix, at the end of the paper, we include
an alternative proof of a statement we used in a discussion
of Step 2.
It is proved in the Appendix that a smooth real-valued function,
which is defined  
on a smooth connected manifold $M$ without boundary,
and  has no zeros of infinite order on $M$,
is uniquely, up to a sign, determined on $M$ by its square.
This proof was communicated to me on behalf of       
Professor Yu. M. Berezansky.

This completes the description of Step 2.

The $C^{\infty}$ smoothness of the data is assumed for technical
reasons: it guarantees the existence of an integer $m$ in the above
construction. 
It would be interesting to weaken this assumption and to find out if 
one can prove Lemma 3.1 assuming $S$ is $C^{1,1}-$smooth
and $q\in L^{\infty}(D)$. 

In section 2 we sketch for
convenience of the reader  a proof of Theorem 1.1.
In section 3 a detailed discussion of Step 2 is given.

\section{Proof of Theorem 1.1} 

The argument of this section is based on the ideas from (\cite{R3})

Without loss of generality assume that $\lambda_j \neq 0$ $\forall j$
and take in what follows $n=3$. The argument for $n>3$
is essentially the same. Let

\begin{equation}
Lu:=-\Delta u + q(x) u = 0 \hbox{\ in\ } D, \quad u_N = f \hbox{\ on\ } S.
\tag{2.1}\end{equation}

Then
\begin{equation}
u(x) = \sum^\infty_{j=1} \frac{\int_S f(s) \varphi_j (s) ds}{\lambda_j}
\varphi_j(x).
\tag{2.2}\end{equation}

Assume that $q_1(x)$ and $q_2 (x)$ generate the same data
\begin{equation}
\{\lambda_j, \varphi_j (s) \}_{\forall j}.
\tag{2.3}\end{equation}

Take an arbitrary $f \in H^{\frac{3}{2}} (S)$, denote
$p(x) := q_2(x) - q_1 (x), \quad w(x) := u_1 (x) - u_2 (x)$,
where $u_m (x)$ is the solution to (2.1) with $q=q_m(x)$, and subtract from
equation (1.1) with $q=q_1$ equation (1.1) with $q=q_2$. The result is

\begin{equation}
L_1 w := -\Delta w + q_1 w = p u_2, \quad \hbox{\ on\ } S,
\tag{2.4}\end{equation}
where $L_m:= -\nabla^2 +q_m(x)$, $m=1,2.$ Clearly $w_N=0$ on $S$.
We will prove later that
\begin{equation}
w=0 \hbox{\ on\ } S.
\tag{2.5}\end{equation}

Assuming (2.5), multiply (2.4) by an arbitrary
$v_1 \in N(L_1) := \{ v_1 : L_1v = 0 \hbox{\ in\ } D, v_1 \in H^2
(D) \}$,
where $H^2(D)$ is the Sobolev space, integrate by parts, use the boundary
conditions (2.4) and (2.5) and get the orthogonality relation:

\begin{equation}
0 = \int_D p(x) u_2 (x) v_1 (x) dx \quad \forall v_1 \in N (L_1),
\quad \forall u_2 \in N(L_2).
\tag{2.6}\end{equation}

By property $C$ ( see \cite{R1}), (2.6) implies $p(x) = 0$, i.e.
$q_1(x)=q_2(x)$.

Recall that property $C$ means that the set of products
$\{ v_1 (x) u_2(x) \}_{\forall v_1 \in N(L_1) \forall u_2 \in N(L_2)}$
is complete in $L^p(D), p \geq 1$. A detailed discussion and proof of  this
property is given in \cite{R1}.

To complete the proof of Theorem 1.1, note that 
(2.5)
follows from (2.2) and from the basic assumption that $q_1$ and $q_2$
generate the same data (2.3). From (2.2) it follows formally that
$u_1(s)=u_2(s)$. Indeed, if $\varphi_j^{(1)} (s) = \varphi_j^{(2)}(s)$
and $\lambda_j^{(1)} = \lambda_j^{(2)} \quad \forall j$, where
$\varphi_j^{(m)} \hbox{\ and\ } \lambda_j^{(m)}$ are the eigenfunctions
and eigenvalues of
the operator $L_m, m=1,2, \hbox{\ with\ } q=q_m$, then formula (2.2) with
$x=s \in S$ shows that $u_1=u_2 \hbox{\ on\ }S$.

However, the series (2.2) converges in $L^2(D)$ but not necessarily on $S$.
Therefore a justification of the above argument is needed. 
Such a justification can be given 
folllowing an idea in [10]. 
Theorem 1.1 is proved. \qed

\section{Finding $\varphi_j(s)$ from $\varphi^2_j (s)$.} 

In the introduction a method for finding $\varphi_j(s)$ from the knowledge
of $\varphi_j^2 (s)$ has been discussed.

Here a justification of this method is presented.
This justification
consists mainly of the proof of the following lemma.

\begin{lemma} 

The function $\varphi_j (s)$ does not have zeros of infinite order.
\end{lemma}

\begin{proof}
It is known (see \cite{H}, p.14, where a stronger
result is obtained) that a solution to the
second order elliptic inequality:
\begin{equation}
|M_0u| \leq c(|u|+| grad u|) \,\, \hbox{\ in\ } D,
\tag{3.1}\end{equation}
where $c=const>0,$ and $M_0$ is a strictly elliptic homogeneous second
order differential expression (summation is understood
over the repeated indices):
\begin{equation}
M_0u := -g_{mj}u_{mj} , 
\quad u_{mj} := \frac{\partial^2 u}{\partial x_m \, \partial x_j},
\tag{3.2}\end{equation}
cannot have a zero of infinite order at a point $y\in D$ provided that:
\begin{equation}
g_{mj} (x) \hbox{ are Lipschitz-continuous and strictly
elliptic,}
\tag{3.3}\end{equation}
and
\begin{equation}
g_{mj} (0) \hbox{ are real numbers. }
\tag{3.4}\end{equation}

By zero of infinite order of a solution $u$ of (3.1) a point $y \in D$ is
meant such that
\begin{equation}
\int_{|y-x|< \epsilon} |u(x)|^2 dx \leq c_m \epsilon ^m \quad
\forall \epsilon > 0, \forall m=1,2,....,
\tag{3.5}\end{equation}
where $c_m$ are positive constants independent of $\epsilon$, and
$\epsilon > 0$ is sufficiently small so that the ball
$B(y, \epsilon) := \{x: |x-y| \leq \epsilon \} \subset D$.

We use this result to prove that the same is true if
$S \in C^{1,1} \hbox{\ and\ } y \in S$.
Let $x_n = f(x^\prime), x^\prime =(x_1, \dots ,x_{n-1})$, be the 
local equation of $S$
in a neighborhood of the point $y \in S,$ which we choose as the 
origin, and
$x_n$-axis be directed along the normal to $S$ pointed into
$D^\prime := \R^n \backslash D$.

Let us introduce the new orthogonal coordinates
$$\xi_j = \xi_j(x_i),\quad 1 \leq j,i \leq n, 
$$
so that $\xi_n = 0$ is the equation of $S$ in a neighborhood of the
origin, and assume $n=3$. 
For example, one can use the coordinate system in
which the $z-$axis is directed along the outer normal to $S$,
and the $x,y-$coordinates are isothermal coordinates on $S$,
which are known to exist for two-dimensional $C^\infty-$surfaces
in $R^3$ ( and even for $C^3-$surfaces in $R^3$, \cite{gu}, p. 246).
For an
arbitrary $C^\infty-$surface  in $R^3$ one can
prove that
locally one can introduce (non-uniquely) the coordinates in which
the metric tensor is diagonal in
a neighborhood of $S$. To do this, one
takes two arbitrary linearly independent vector fields tangent
to $S$ and orthogonalize them with respect to the 
Euclidean metric in $R^3$ using the Gram-Schmidt procedure, which
is always possible. The third axis of the coordinate
system, which we are constructing, is directed along the normal to $S$
at each point of the patch on $S$.
Let $A$ and $B$ be the resulting orthogonal vector fields
tangent to $S$. Then one can find (non-uniquely) a function $\chi$, 
defined on the local chart, such
that the Lie bracket of the vector fields $wA$ and $B$ vanishes, 
$[\chi A, B]=0$. Then the flows of the vector fields $\chi A$
and $B$ commute and provide the desired orthogonal coordinate
system in which the metric tensor is diagonal (\cite{cm}).
The condition $[\chi A, B]=0$ can be written as
the following linear partial differential equation:
$\chi h_x-h\chi_x-\chi_y=0$, where $h:=-\frac {f_xf_y}{1+f_x^2}$,
$x$ and $y$ are the parameters, and $z=f(x,y)$ is the local
equation of the surface $S$ on the chart. The above equation for $\chi$
has many solutions. One can find a solution $\chi (x,y)$ by the
standard method of characteristics. A unique solution is specified by
prescribing some Cauchy data, which geometrically means that
a noncharacteristic curve through which the surface
$\chi=\chi (x,y)$ passes, should be specified (\cite{cg}). 

  If $n>3$ the situation is less simple if one wants
to use the same idea in the argument: there are many
($(n-1)(n-2)/2$) Lie brackets to vanish in the case of $n-1$
vector fields tangent to $S$ in $R^n$, and it is 
not clear for what $S$
these conditions can be satisfied. However, for our argument
it is sufficient to have the coordinate system in which
the metric tensor has zero elements $g^{jn}$ for $1\leq j \leq n-1$,
and $g^{nn}$ does not vanish. In this case the even
continuation (3.8)-(3.10), that is used below, still allows one
to claim that the function (3.8) solves the same equation in
the region $\xi_n>0$ as it solves in the original region  $\xi_n<0$.
If the elements $g^{jn}$ for $1\leq j \leq n-1$ 
do not vanish, then the
equation in the region $\xi_n>0$ will have some of the coefficients in
front of the second mixed 
derivatives $g^{jn}w_{jn}$ with the minus sign,
while these coefficients in the region
 $\xi_n<0$ enter with the plus sign.
So, in this case the principal part $M_0$ of the operator $M$,
which is used in (3.11), will be different in the regions
$\xi_n>0$ and  $\xi_n<0$. Recall that the Laplace operator in the new
coordinates has the form of the Laplace-Beltrami
operator: $\Delta w=g^{-1/2}(g^{1/2} g^{ij}w_j)_i$, where $g:=det
(g_{ij})$,
summation is understood over the repeated indices, and 
$w_j:=\frac {\partial w}{\partial \xi_j}$.

 If $y\in S$ is a zero of $\varphi_j(s)$ of infinite
order
in the sense $|\varphi_j(s)|\leq c_m |s-y|^m \quad \forall m=1,2,....,
s\in S$, then equations (1.1) and (1.2) imply that 
$|\varphi_j(x)|\leq c_m |x-y|^m \quad \forall m=1,2,....,
x\in D \cap B (y, \epsilon)$, so that condition
similar to (3.5) holds for the integrals 
over $D \cap B (y, \epsilon)$.
 Indeed, the derivatives of $\varphi_j(x)$ in the tangential to $S$
directions at the point $y\in S$ vanish by the assumption.
From (1.1) and (1.2) it follows that the normal derivatives
of $\varphi_j(x)$ of the first and second order vanish at $y$.
Differentiating equation (1.1) along the normal one concludes that all
the normal derivatives of $\varphi_j(x)$ vanish at the point $y$.
Thus we may assume 
that $y\in S$ is a zero of $\varphi_j(x)$ of infinite
order, that is, inequalities  (3.5)
hold for $u = \varphi_j(x)$ for 
the integral over $D \cap B (y, \epsilon)$.

In the $\xi$- coordinates one writes the equation for $\varphi_j$:

\begin{equation}
-\Delta \varphi_j + q(x) \varphi_j - \lambda_j \varphi_j (x) = 0
\hbox{\ in\ } D, \quad \varphi_{jN} = 0 \hbox{\ on\ } S,
\tag{3.6}\end{equation}
as follows:
\begin{equation}
M \varphi_j = 0 \hbox{\ for\ } \xi_n <0 \hbox{\ in\ } D, \quad
\varphi_{j \xi_n} = 0 \hbox{\ at\ } \xi_n = 0.
\tag{3.7}\end{equation}

We drop the subscript $j$ of $\varphi_j$ and of $\lambda_j$ in what
follows.
In (3.7) the operator $M$ is defined as:

$$M \varphi =- \frac{1}{g_{jj}}
\frac{\partial^2 \varphi}{\partial \xi_j^2}- (\frac{1}{g^{1/2}}
\frac{\partial (g_{jj}^{-1}g^{1/2}) }{\partial \xi_j}\frac{\partial
\varphi}{\partial \xi_j}) + Q(\xi) \varphi -
\lambda \varphi,
\quad Q(\xi) := q(x(\xi)),$$
over the repeated indices summation is understood,
$g_{ij}$ is the metric tensor of the new coordinate system,
$g:=det (g_{ij}),$ and the
coefficients
in front of the second-order derivatives in $M$ are
extended to the region $\xi_n>0$ as even functions of $\xi_n$,
so that the extended coefficients are Lipschitz in a ball
centered at $y\in S$ with radius $\epsilon>0.$

Let us define $w$ in a neighborhood of the origin, $|\xi| < \epsilon$,
by setting
\begin{equation}
w=\left\{ \begin{array}{ll}
\varphi(\xi',\xi_n) & \hbox{\ if\ } \xi_n \leq 0,\\
\varphi(\xi',-\xi_n) & \hbox{\ if\ } \xi_n>0, 
\end{array}\right.
\tag{3.8}\end{equation}
and 
\begin{equation}
Q(\xi)=\left\{ \begin{array}{ll}
Q(\xi',\xi_n) & \hbox{\ if\ } \xi_n \leq 0,\\
Q(\xi',-\xi_n) & \hbox{\ if\ } \xi_n>0,
\end{array} \right.
\tag{3.9}\end{equation}
\begin{equation}
g_{jj}(\xi)=\left\{ \begin{array}{ll}
g_{jj}(\xi', \xi_n) & \hbox{\ if\ } \xi_n\leq 0,\\
g_{jj}(\xi',-\xi_n) & \hbox{\ if\ } \xi_n >0.
\end{array} \right.
\tag{3.10}\end{equation}

The functions $g_{jj} (\xi),$ defined by (3.10), are 
Lipschitz in the ball $B(0, \epsilon)$,  if
$g_{jj}(\xi)$ is Lipschitz in
$B(0, \epsilon) \cap R^n_-$, where $R^n_- := \{ \xi : \xi_n \leq 0 \}$.

Furthermore,
\begin{equation}
|M_0w|\leq c(|w|+| grad w|)  \hbox{\ in\ } B(0, \epsilon),
\tag{3.11}\end{equation}
where $c=const>0,$ and
\begin{equation}
\int_{B(0, \epsilon)} |w|^2d \xi \leq \tilde c_m \epsilon^m 
\quad \forall m=1,2,.....,
\tag{3.12}\end{equation}
if one assumes
\begin{equation}
\int_{B(0, \epsilon) \cap \R_-^n} |\varphi|^2 d \xi 
\leq  \tilde c_m \epsilon^m \quad \forall m=1,2,......,
\tag{3.13}\end{equation}
for all sufficiently small $\epsilon>0$.

Note that the change of variables $x\to \xi$ is a
smooth diffeomorphism in a neighborhood of the origin,
which maps the region $D\cap B(0, \epsilon)$ onto
a neighborhood of the origin in $ \R_-^n$ in $\xi-$ space,
and one can always choose a half-ball $B(0, \epsilon)$ belonging to this
neighborhood, so that (3.13) follows from (3.5).
 
Therefore $w \equiv 0$ in $B(0, \epsilon)$ and consequently 
$\varphi \equiv 0$ in
$B(0, \epsilon) \cap \R^n_-$. This implies that $\varphi \equiv 0$ in $D$
by the unique continuation theorem (see \cite{H} and \cite{W}).
This is a contradiction since $||\varphi||=1$. 

Lemma 3.1 is proved.
$\Box$

This lemma provides a justification of the argument given for Step 2 in the
introduction.

\begin{remark}  

In this remark we comment on the numerical recovery of the potential from the
data $\theta (s,s,\lambda)$. We have explained how to get the data
$\{\lambda_j, \varphi_j (s) \}_{\forall j}$ from $\theta (s,s, \lambda)$.
Therefore the spectral function
\begin{equation}
\theta (s,t, \lambda) = \sum_{\lambda_j < \lambda} \varphi_j (s)
\varphi_j (t), \quad \forall s, t \in S
\tag{3.14} \end{equation}
and the resolvent kernel
\begin{equation}
G(s,t,\lambda) = \int^\infty_{-\infty} \frac{d_{\mu} \theta
(s,t,\mu)}{\mu-\lambda}
\tag{3.15}\end{equation}
are known on $S$. If $G(s,t,\lambda)$ is known, then the Neumann-to-Dirichlet
(N-D) map $u_N := f \to h := u|_S$ is known. This map $\Lambda : f \to h$
at every fixed $\lambda \neq \lambda_j$ $\forall j,$ associates to every
$f \in H^{\frac{1}{2}} (S)$ a function $h \in H^{\frac{3}{2}} (S)$
by the formula
\begin{equation}
h(s) = \int_S G (s,t,\mu) f (t) dt.
\tag{3.16}\end{equation}
Here
\begin{equation}
Lu - \lambda u = 0 \hbox{\ in\ } D, \quad u_N = f \hbox{\ on\ }S.
\tag{3.17}\end{equation}

The unique solution to (3.17) is given by (3.16).
\end{remark}

We now outline an alternative proof of Theorem 1.2:
steps 1 and 2 are the same as in Theorem 1.2; step 3
consists of the construction of the N-D map by
formula (3.16) and reference to Theorem 3.1 below.

In \cite{R1} and \cite{R4} 
the result similar to Theorem 3.1 (see below) is proved for the 
D-N (Dirichlet-to-Neumann) map. 
Our proof follows the idea of the proof in
\cite{R4}.
As above, $L_m, m=1,2,$ stands for the Neumann operator
$-\Delta + q_m  \hbox{\ in\ } D$. Let us assume that $ \lambda \in R$
is a regular point for $L_1$ and $L_2$.

\begin{theorem} 

If $L_1 -\lambda$ and $L_2 -\lambda$ generate the same N-D map then
$q_1=q_2$.
\end{theorem}

\begin{proof}
Take an arbitrary $f \in H^{\frac{1}{2}} (S)$ and consider the problems

$ (L_m - \lambda) u_m = 0 \hbox{\ in\ } D, u_{mN} = f \hbox{\ on\ }S$.

Subtract from the equation with $m=1$ the equation with $m=2$ and get
\begin{equation}
(L_m - \lambda) w=pu_2 \hbox{\ in\ } D, \quad w_N = w = 0 \hbox{\ on\ } S,
\tag{3.18}\end{equation}
where $w := u_1-u_2, p := q_2-q_1$. The condition $w=0$ on $S$
follows from the basic assumption,
namely the N-D map is the same for $L_1$ and
$L_2$.
Let $\psi_1 \in N (L_1) := \{\psi : L_1 \psi = 0 \hbox{\ in\ }D,
\psi \in H^2(D)\}$ be arbitrary. Multiply (3.18) by $\psi_1,$ integrate by
parts using boundary conditions (3.18), and get the orthogonality
relation:
\begin{equation}
\int_D p u_2 \psi_1 dx = 0 \quad \forall \psi_1 \in N(L_1) \quad
\forall u_2 \in N (L_2).
\tag{3.19}\end{equation}

The last inclusion in (3.19) follows since $f \in H^{\frac{1}{2}} (S)$
is arbitrary. From (3.19) and property $C$ (see \cite{R1}) it follows that
$p(x) \equiv 0$. Theorem 3.1 is proved.   \end{proof}

In \cite{R1} \cite{R5}--\cite{R6} one finds a discussion
of the inversion algorithm.

\nd{\bf Appendix }

 {\bf Lemma}.  {\it Let $M$ be a connected $C^{\infty}$ manifold
and suppose that a real-valued function $f\in C^{\infty}(M)$ does not
have
zeros of infinite order, i.e., for an arbitrary point $x_0\in M$,
$f^{(\alpha)}(x_0)\not =0$ for some $\alpha$. Then, up to sign, the
function $f$ is uniquely defined on $M$ by its square $f^2$.}

{\it Proof.} Let $g\in C^{\infty}(M)$ be a real-valued function
satisfying the condition $g^2=f^2$. Then the lemma states that either
$g=f$ or $g=-f$. Consider the function $u=f-g$. If $u\equiv 0$ on $M$,
then $g\equiv f$ and the lemma is proved. Consider $X=\{x:u(x)\not
=0,x\in M\}$.
This is an open set and $u(x)=2f(x)$ on this set, since, for a fixed
$x$, the
condition $g^2(x)=f^2(x)$ implies that  $f(x)=g(x)$ or  $g(x)=-f(x)$.
The equality $u(x)=2f(x)$ will also hold on the closure
$\overline X$.
If
$\overline X=M$, then $u=f-g=2f$, i.e., $g\equiv -f$ on $M$, and
the
lemma is proved. Let $Y=M\setminus \overline X$ be a nonempty 
open set.
Since
$M=\overline X\cup \overline Y$ and $M$ is connected, there exists a
point
$x_0\in \overline X\cap \overline Y$. 
We have $u(x)=2f(x)$ on $\overline X$,
and the function $f$ does not have a zero of infinite order, 
hence there
exists an $\alpha$ such that
$u^{(\alpha)}(x_0)=2f^{(\alpha)}(x_0)\not =
0$.
On the other hand, $x_0\in \overline Y$ and $u(x)\equiv 0$ on the open
set $Y$.
Thus $u^{(\alpha)}(y) =0$ for all $y\in Y$. This and
the continuity of $u^{\alpha}(\cdot)$ imply, as $y\to x_0$, that 
$u^{(\alpha)}(x_0)=0$. This contradiction proves the lemma.
\end{proof}
\medskip


\nd{\bf Acknowledgement}

The authors thank Professors Yu. M. Berezansky, D. Jerison,
 F. Gesztesy, S. Kuzhel and F. Miller for useful remarks and discussions. 
This paper was written when the author was 
supported by a big 12 fellowship and visited
the University of Missouri at Columbia.
The author thanks this University for hospitality.
\pagebreak

\end{document}